\documentclass[conference]{IEEEtran}
\IEEEoverridecommandlockouts
\usepackage{cite}
\usepackage{amsmath,amssymb,amsfonts}
\usepackage{graphicx}
\usepackage{multirow,array,booktabs}
\usepackage[table]{xcolor}
\usepackage{tikz}
\usetikzlibrary{positioning,arrows.meta,fit}
\usepackage{url}
\usepackage{balance}
\usepackage{hyperref}
\hypersetup{hidelinks}

\definecolor{resultgreen}{RGB}{228,242,230}
\definecolor{resultred}{RGB}{251,229,229}
\newcommand{\favorable}[1]{\cellcolor{resultgreen}{\bfseries\boldmath #1}}
\newcommand{\tradeoff}[1]{\cellcolor{resultred}{\bfseries\boldmath #1}}

\newcommand{\keyresult}[2]{%
  \par\smallskip\noindent
  \colorbox{black!8}{\footnotesize\bfseries Finding~#1}%
  \enspace\textbf{#2}%
}

\begin{document}
\title{Engineering Low-Latency, High-Throughput Quantum Decoders on NVIDIA GPUs}
\author{\IEEEauthorblockN{Stefano Markidis, Luca Pennati, Ivy Peng}
\IEEEauthorblockA{KTH Royal Institute of Technology}}
\maketitle

\begin{abstract}
We design and optimize a quantum error correction~(QEC) decoder for NVIDIA GPUs, combining Relay belief propagation, a state-of-the-art message-passing algorithm, with order-zero ordered statistics post-processing. Our implementation uses CUDA Graphs to control GPU execution, stops updates for converged samples, and collects unresolved samples in queues for post-processing. We optimize sparse traversal through coalesced message accesses, prefetching, and register reuse, and accelerate binary post-processing with warp-level collectives. In addition, mixed FP16–FP32 arithmetic and INT8 quantization reduce message storage and traffic while retaining accuracy using accumulators at higher precision. We evaluate surface, hypergraph product, and bivariate bicycle code workloads on NVIDIA A100 and GH200 data-center GPUs and a commodity RTX 5080. On the bivariate bicycle code at error probability 0.003, INT8 halves amortized decoding time relative to independently tuned FP32 with a negligible logical error rate increase on RTX~5080. With independent syndrome batches submitted back-to-back to concurrent decoder instances, RTX~5080 delivers 91\% of GH200's highest measured throughput with an 8.9$\times$ shorter median submission-to-correction latency. The combination of performance optimization, mixed precision, INT8 quantization, and concurrent decoder instances opens a path toward QEC decoder clusters built from commodity GPUs.
\end{abstract} 


\section{Introduction}
Fault-tolerant quantum computing requires classical decoders that process measurement data at least as fast as it is generated~\cite{GoogleQuantumAI2025WillowQEC}. Quantum error correction~(QEC) encodes one or more logical qubits in a \emph{code block} of physical qubits. Repeated parity measurements, called \emph{stabilizer checks}, produce a \emph{syndrome}: a binary vector recording deviations from expected correct outcomes. The decoder infers a correction from this vector. As the number of logical qubits (each encoded across multiple physical
qubits) grows, each measurement round generates more syndrome data, increasing the required aggregate decoding throughput. Independent decoding requests provide parallel work that the classical system can exploit to sustain this higher data rate.

From an HPC perspective, the QEC computational workload is akin to sparse graph processing. QEC graphs have nodes with different numbers of neighbors. Each edge update requires few arithmetic operations, and iteration counts depend on the measured syndrome. Reconfigurable FPGAs and custom ASICs can tailor their logic and memory organization to these operations~\cite{Bascones2025FPGAASICBPOSD}. GPUs combine high memory bandwidth with thread parallelism across graph updates and independent syndrome samples, allowing memory accesses to overlap. In addition, CUDA provides the programming model, libraries, and profiling tools for specializing kernels and controlling concurrent execution. In this work, we evaluate the latency, throughput, and logical accuracy of optimized GPU decoding to assess whether GPUs can serve as credible, programmable devices for QEC workloads.

\begin{figure}[t]
\centering
\includegraphics[width=0.7\columnwidth]{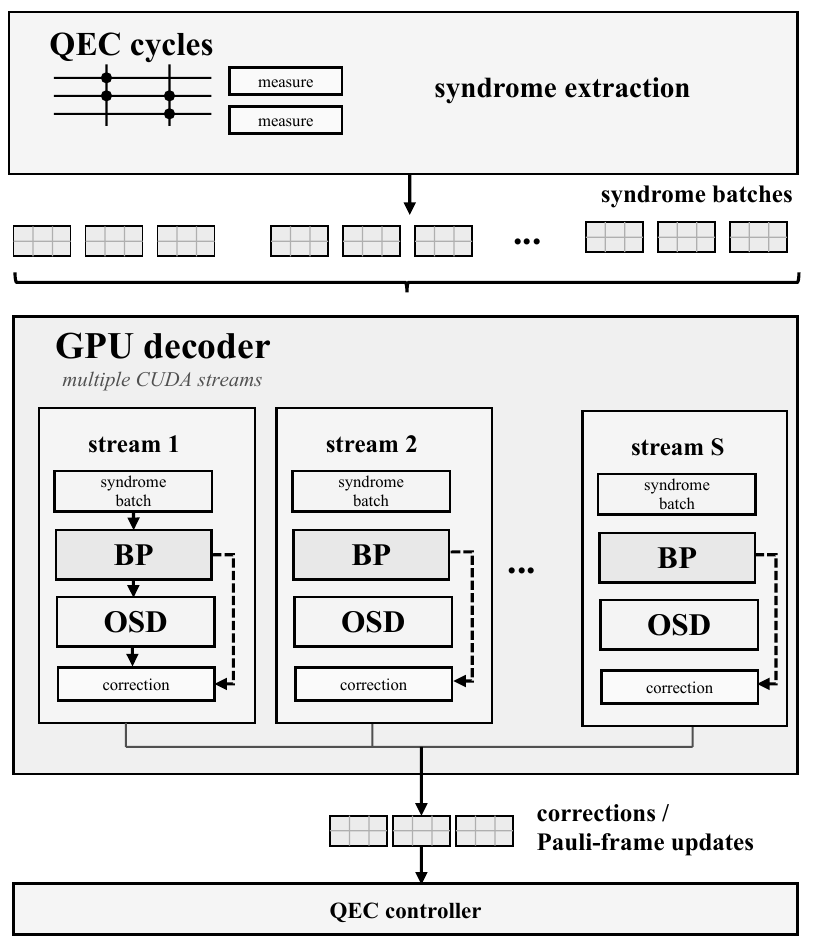}
\caption{GPU decoding workflow in a QEC system.}
\label{fig:multistream_qec}
\end{figure}
QEC decoding exposes GPU parallelism within each syndrome sample and across independent decoding requests. Each sample requires repeated sparse graph traversal, followed by a binary linear solver if the iterative decoder reaches its iteration limit without converging. GPU threads distribute graph updates and cooperate on local reductions and binary operations. Samples share the same graph connectivity but maintain independent messages and convergence state, allowing many samples to be processed together. Fig.~\ref{fig:multistream_qec} shows how this computation fits into the overall workflow: independent syndrome batches execute in separate CUDA streams and return corrections to the controller. The controller updates a \emph{Pauli frame}, a classical record of accumulated bit and phase corrections. Our work develops this mapping for Relay belief propagation~(Relay-BP) with order-zero ordered statistics post-processing (OSD-0), combining parallelism within each decoder instance with concurrency across batches.

We design and optimize a Relay-BP/OSD-0 decoder for NVIDIA GPUs, targeting high throughput and short correction latency while quantifying the effects of numerical precision on accuracy. The main contributions are:
\begin{itemize}
\item A GPU decoder with convergence control on the GPU, sample retirement, compacted fallback queues, and ordered multi-warp binary elimination.
\item Specialized sparse kernels using mixed FP16-FP32 arithmetic, FP16 arithmetic, and INT8 quantization, with an analysis of their effects
on memory footprint, convergence, and complete-decoder performance.
\item An evaluation across three circuit-level code families and three NVIDIA GPU architectures, combining kernel profiling and latency-throughput measurements to identify precision, batch size, and stream configurations suited to different response-time requirements.
\end{itemize}

\section{A Quantum Error Correction Workflow on NVIDIA GPUs}
\label{sec:workflow} 
This section introduces the QEC terminology and decoding algorithms needed to develop the GPU decoder. We begin with the sparse graph representation of circuit-level QEC workloads, then describe Relay-BP message passing and the OSD-0 binary solver.

\begin{figure}[t]
\centering
\resizebox{\columnwidth}{!}{
\begingroup
\fontfamily{ptm}\selectfont
\begin{tikzpicture}[x=1cm,y=1cm,font=\scriptsize,>=Latex]
\node[font=\scriptsize\bfseries] at (2.30,3.02) {(a) Sparse check matrix};
\node[font=\scriptsize\bfseries] at (6.32,3.02) {(b) Tanner subgraph};
\begin{scope}[shift={(0.55,2.65)},x=0.0426829268cm,y=-0.0560975610cm]
\foreach \x in {0.5,1.5,...,81.5}
  \draw[black!18,line width=0.12pt] (\x,0.5) -- (\x,40.5);
\foreach \y in {0.5,1.5,...,40.5}
  \draw[black!18,line width=0.12pt] (0.5,\y) -- (81.5,\y);
\foreach \r/\v in {
  0/0,0/1,0/9,0/10,
  1/2,1/3,1/11,1/12,
  2/4,2/5,2/13,2/14,
  3/6,3/7,3/15,3/16,
  4/10,4/11,4/19,4/20,
  5/12,5/13,5/21,5/22,
  6/14,6/15,6/23,6/24,
  7/16,7/17,7/25,7/26,
  8/18,8/19,8/27,8/28,
  9/20,9/21,9/29,9/30,
  10/22,10/23,10/31,10/32,
  11/24,11/25,11/33,11/34,
  12/28,12/29,12/37,12/38,
  13/30,13/31,13/39,13/40,
  14/32,14/33,14/41,14/42,
  15/34,15/35,15/43,15/44,
  16/36,16/37,16/45,16/46,
  17/38,17/39,17/47,17/48,
  18/40,18/41,18/49,18/50,
  19/42,19/43,19/51,19/52,
  20/46,20/47,20/55,20/56,
  21/48,21/49,21/57,21/58,
  22/50,22/51,22/59,22/60,
  23/52,23/53,23/61,23/62,
  24/54,24/55,24/63,24/64,
  25/56,25/57,25/65,25/66,
  26/58,26/59,26/67,26/68,
  27/60,27/61,27/69,27/70,
  28/64,28/65,28/73,28/74,
  29/66,29/67,29/75,29/76,
  30/68,30/69,30/77,30/78,
  31/70,31/71,31/79,31/80,
  32/1,32/2,
  33/3,33/4,
  34/5,34/6,
  35/7,35/8,
  36/72,36/73,
  37/74,37/75,
  38/76,38/77,
  39/78,39/79} {
  \fill[black!65] ({\v+0.58},{\r+0.62}) rectangle ({\v+1.42},{\r+1.38});
}
\fill[black] (15.58,3.62) rectangle (16.42,4.38);
\end{scope}
\draw[black,line width=0.45pt] (0.55,0.35) rectangle (4.05,2.65);
\foreach \t in {0,20,40,60,80} {
  \pgfmathsetmacro{\xt}{0.55+3.5*(\t+1)/82}
  \draw[line width=0.4pt] (\xt,0.35) -- ++(0,0.045);
  \node[below,inner sep=1pt] at (\xt,0.33) {\t};
}
\foreach \t in {0,10,20,30,40} {
  \pgfmathsetmacro{\yt}{2.65-2.3*(\t+1)/41}
  \draw[line width=0.4pt] (0.55,\yt) -- ++(0.045,0);
  \node[left,inner sep=1pt] at (0.53,\yt) {\t};
}
\node at (2.30,-0.12) {variable index \textit{v}};
\node[rotate=90] at (-0.06,1.50) {check index \textit{c}};
\node at (4.96,2.73) {checks};
\node at (7.11,2.73) {variables};
\node[draw,rectangle,fill=black!15,minimum size=3.2mm,inner sep=0pt] (c3) at (4.96,2.32) {};
\node[left=1mm of c3] {\textit{c}\textsubscript{3}};
\node[draw,rectangle,fill=white,minimum size=3.2mm,inner sep=0pt] (c6) at (4.96,1.48) {};
\node[left=1mm of c6] {\textit{c}\textsubscript{6}};
\node[draw,rectangle,fill=white,minimum size=3.2mm,inner sep=0pt] (c7) at (4.96,0.56) {};
\node[left=1mm of c7] {\textit{c}\textsubscript{7}};
\node[draw,circle,fill=white,minimum size=2.1mm,inner sep=0pt] (v6) at (7.00,2.48000) {};
\node[right=0.7mm of v6] {\textit{v}\textsubscript{6}};
\node[draw,circle,fill=white,minimum size=2.1mm,inner sep=0pt] (v7) at (7.00,2.23778) {};
\node[right=0.7mm of v7] {\textit{v}\textsubscript{7}};
\node[draw,circle,fill=white,minimum size=2.1mm,inner sep=0pt] (v14) at (7.00,1.99556) {};
\node[right=0.7mm of v14] {\textit{v}\textsubscript{14}};
\node[draw,circle,fill=black!15,minimum size=2.1mm,inner sep=0pt] (v15) at (7.00,1.75333) {};
\node[right=0.7mm of v15] {\textit{v}\textsubscript{15}};
\node[draw,circle,fill=white,minimum size=2.1mm,inner sep=0pt] (v16) at (7.00,1.51111) {};
\node[right=0.7mm of v16] {\textit{v}\textsubscript{16}};
\node[draw,circle,fill=white,minimum size=2.1mm,inner sep=0pt] (v17) at (7.00,1.26889) {};
\node[right=0.7mm of v17] {\textit{v}\textsubscript{17}};
\node[draw,circle,fill=white,minimum size=2.1mm,inner sep=0pt] (v23) at (7.00,1.02667) {};
\node[right=0.7mm of v23] {\textit{v}\textsubscript{23}};
\node[draw,circle,fill=white,minimum size=2.1mm,inner sep=0pt] (v24) at (7.00,0.78444) {};
\node[right=0.7mm of v24] {\textit{v}\textsubscript{24}};
\node[draw,circle,fill=white,minimum size=2.1mm,inner sep=0pt] (v25) at (7.00,0.54222) {};
\node[right=0.7mm of v25] {\textit{v}\textsubscript{25}};
\node[draw,circle,fill=white,minimum size=2.1mm,inner sep=0pt] (v26) at (7.00,0.30000) {};
\node[right=0.7mm of v26] {\textit{v}\textsubscript{26}};
\draw[black!45,line width=0.35pt] (c3.east) -- (v6.west);
\draw[black!45,line width=0.35pt] (c3.east) -- (v7.west);
\draw[black!45,line width=0.35pt] (c3.east) -- (v16.west);
\draw[black!45,line width=0.35pt] (c6.east) -- (v14.west);
\draw[black!45,line width=0.35pt] (c6.east) -- (v15.west);
\draw[black!45,line width=0.35pt] (c6.east) -- (v23.west);
\draw[black!45,line width=0.35pt] (c6.east) -- (v24.west);
\draw[black!45,line width=0.35pt] (c7.east) -- (v16.west);
\draw[black!45,line width=0.35pt] (c7.east) -- (v17.west);
\draw[black!45,line width=0.35pt] (c7.east) -- (v25.west);
\draw[black!45,line width=0.35pt] (c7.east) -- (v26.west);
\draw[->,black,semithick] (c3.east) to[bend left=9] (v15.west);
\draw[->,black,semithick,dash pattern=on 2pt off 1.3pt] (v15.west) to[bend left=9] (c3.east);
\node[align=center] at (6.20,-0.12) {two directed messages per edge};
\end{tikzpicture}
\endgroup}
\caption{Sparse check matrix (a) and the corresponding Tanner subgraph (b). Each nonzero defines a check-variable edge.}
\label{fig:sparse_graph}
\end{figure}
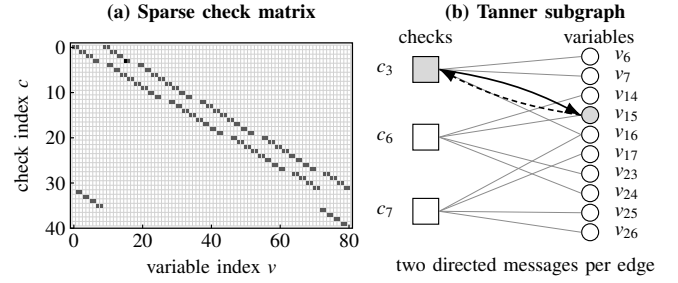

\noindent\textbf{QEC Sparse Matrices and Decoding Graphs.} The core data structure in graph-based QEC decoding is a sparse binary check matrix $H$. It records how possible errors affect measured check outcomes: each column represents an error variable and each row a parity check. An entry is one if that error flips the corresponding check outcome, and zero otherwise. The matrix defines a bipartite Tanner graph, with separate node sets for checks and variables and one edge per nonzero (see Fig.~\ref{fig:sparse_graph}). Row and column traversals access the neighborhoods of checks and variables, respectively.

To infer a correction from syndrome data, the decoder needs a physical noise model specifying possible errors and their probabilities. Examples include depolarizing noise, which introduces random bit flips, phase flips, or both, and measurement errors, which flip recorded bits. In \emph{circuit-level decoding} (the model we use in this work), this model assigns error probabilities to individual gates, initialization, idle intervals, and measurements in the circuit used to repeatedly measure stabilizer checks. A \emph{detector} produces a bit indicating whether the parity of selected measurement outcomes differs from its expected error-free value. QEC simulation tools such as \texttt{\small{Stim}}~\cite{Gidney2021Stim} can generate syndrome samples from noisy circuits and derive a detector error model~(DEM). The DEM records each error mechanism's probability and the detectors it flips, providing the sparse matrix $H$ and the error-probability vector used by the decoder.

Each circuit execution produces one syndrome sample $\mathbf s$. The decoder seeks a binary error pattern $\hat{\mathbf e}$ satisfying $H\hat{\mathbf e}=\mathbf s\pmod 2$. Each row evaluates an XOR over its incident error bits, which must match the corresponding syndrome bit. The error probabilities guide the choice among compatible patterns. A batch presents multiple independent right-hand sides for the same sparse matrix. During iterative decoding, row and column traversals gather values from check and variable neighborhoods. Their lengths determine the local reduction work, while irregular connectivity determines the indexed memory accesses. Each full graph sweep processes the nonzeros, so its work and per-sample edge storage scale with the edge count. The matrix and error probabilities are reused across the batch. 

\noindent\textbf{Message-Passing QEC Decoders.} Message-passing decoders infer error patterns through repeated local updates on the decoding graph. The computational pattern is similar to neighborhood aggregation in graph neural networks: nodes gather messages from neighbors, combine them, and send updated information along graph edges (Fig.~\ref{fig:sparse_graph}). In belief propagation~(BP), messages represent error reliabilities as log-likelihood ratios. Positive values favor no error, negative values favor an error, and larger magnitudes indicate greater confidence. The \emph{channel prior} is the initial reliability derived from each variable's error probability. 

In this work, we use Relay-BP, which adds memory to BP updates~\cite{Mueller2025RelayBP}. Each iteration executes two dependent kernels: \emph{(i)} \texttt{\small{Parity/min-sum}} gathers variable-to-check messages over each check neighborhood and generates check-to-variable messages. For each check-to-variable message, we exclude the incoming message from the recipient variable. The outgoing sign bit is the XOR of the syndrome bit and the remaining incoming sign bits. We use the smallest remaining magnitude as the outgoing confidence: the least certain neighbor limits the strength of the check's inference. \emph{Min-sum} approximates BP's nonlinear check update with sign and minimum reductions. Normalization scales the minimum magnitude down to compensate for the approximation's overestimate of message confidence. \emph{(ii)} In \texttt{\small{Belief/Relay}}, we form a \emph{posterior}, the updated reliability estimate for each error variable. We combine a weighted blend of the channel prior and previous posterior with the sum of incoming check messages. Subtracting the recipient check's incoming contribution gives the outgoing variable-to-check message. Relay-BP groups iterations into \emph{legs}, each with an iteration limit. A new leg restarts message passing from the channel prior and initializes Relay memory with the preceding leg's final posterior. After each iteration, we make a \emph{hard decision} by converting each posterior into a bit: 1 for a predicted error (negative posterior) and 0 otherwise. We retire a sample, stopping further updates, when this candidate reproduces its measured syndrome. Samples unresolved after the final leg proceed to OSD-0.

\noindent\textbf{OSD-0: A Binary Linear Solver.} Samples unresolved by message passing are passed to OSD-0~\cite{Roffe2020Landscape}, which addresses the inverse problem defined by $H$: the measured syndrome is known, while the error pattern is unknown. For unresolved samples, we compute the \emph{residual syndrome}, the XOR difference between the measured syndrome and that predicted by the BP candidate. We sort variables by increasing absolute posterior and use Gaussian elimination to select linearly independent columns of $H$ in this order. These columns form a basis in which we express the residual syndrome, identifying which error bits to toggle. This ordering prioritizes adjustments to the least confident BP estimates. Order zero constructs a single candidate correction (higher-order OSD explores additional bit flip combinations at greater computational cost). The solver operates over $\mathrm{GF}(2)$ (the Galois field with two elements, 0 and 1), where addition and subtraction are XOR and multiplication is AND. Binary Gaussian elimination selects nonzero pivots and cancels entries through XOR operations. Packing vectors into machine words allows each instruction to update many binary entries simultaneously, i.e., a 64-bit XOR updates 64 entries. Ordered pivot selection (choosing linearly independent columns from least to most reliable) introduces sequential dependencies, while sample-dependent elimination complicates load balancing on GPUs. Hybrid implementations have consequently retained OSD on the CPU even when message passing runs on the GPU. Efficient GPU execution requires parallelism across packed words and unresolved samples while preserving the reliability order of basis selection. Sorting and elimination also make an OSD-0 invocation substantially more expensive than a message-passing iteration. Resolving more samples through Relay-BP therefore reduces costly post-processing.

\begin{figure*}[t]
\centering
\includegraphics[width=0.8\textwidth]{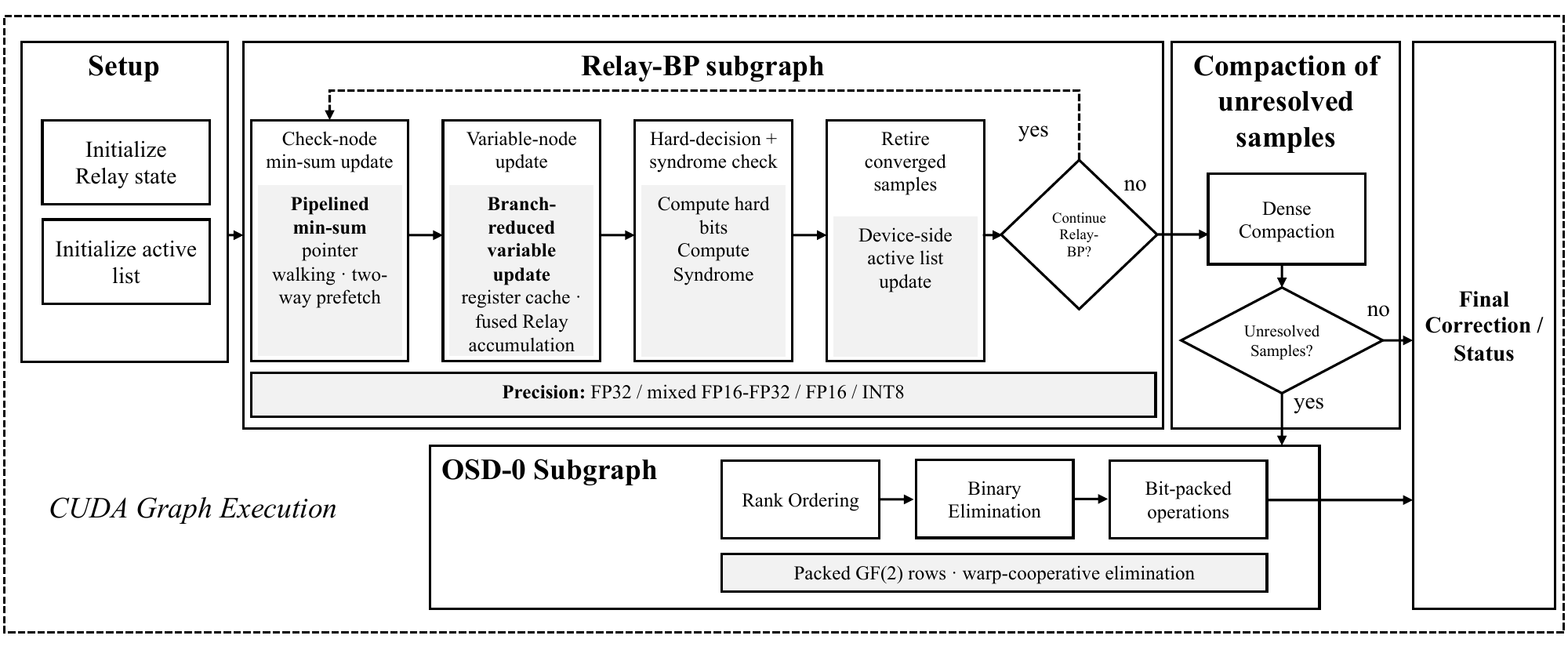}
\caption{GPU workflow within one instantiated CUDA Graph.}
\label{fig:software_structure}
\end{figure*}
\section{Performance Optimization on NVIDIA GPUs}
\label{sec:gpu-optimizations}
This section presents the kernel, numerical, and execution optimizations of the NVIDIA GPU decoder. We follow the execution blocks of the instantiated CUDA Graph in Fig.~\ref{fig:software_structure} to explain sparse traversal, memory reuse, device-side convergence control, and cooperative binary elimination. For the message-passing stages, we detail reduced-precision FP16 arithmetic, mixed FP16-FP32 arithmetic, and INT8 quantization, explaining the choice of message formats, accumulated state, and intermediate precision.

\noindent\textbf{Data Layout and Thread Mapping.} All syndrome samples in a batch share the same decoding graph. We store its connectivity once using compressed sparse row (CSR) lists of variables per check and compressed sparse column (CSC) indexing of incident edge IDs per variable. Maintaining both views gives each update direct neighborhood access, avoiding searches through the full edge list. Both directed message arrays share a common check-major edge numbering and use an edge-major layout, with samples contiguous within each edge. Per-variable state follows the same sample-contiguous ordering. Threads processing adjacent samples traverse the same adjacency list and access contiguous message and state values, coalescing memory accesses across the batch.

The \texttt{\small{Parity/min-sum}} update traverses each check neighborhood twice: first to compute sign parity, the two smallest magnitudes, and the minimum's index, then to generate outgoing messages. In both passes, our optimized variant prefetches two messages into registers before processing either value, software-pipelining loads ahead of dependent arithmetic to increase memory-level parallelism. We advance message pointers by the batch stride to avoid repeated wide edge-offset calculations. In \texttt{\small{Belief/Relay}}, we eliminate repeated loads by retaining incoming messages in registers for variables with at most four neighbors. Each cached value contributes to the posterior sum and is then subtracted to form the corresponding outgoing message. Larger neighborhoods use a general traversal. We further reduce arithmetic and control overhead by precomputing Relay coefficients, using fused multiply-add instructions, and applying branchless clipping.

\noindent\textbf{Convergence and Work Scheduling.} As a further optimization, we encode the dependent iteration kernels in a preinstantiated CUDA Graph and evaluate convergence on the GPU, avoiding host readbacks between iterations. Each Relay leg contains a conditional \texttt{\small{WHILE}} node that repeats the updates while samples remain active and the iteration limit has not been reached. A kernel evaluates this condition using warp ballots and a shared-memory reduction. We retire converged samples by marking their entries inactive in a fixed-capacity sample-ID array. Subsequent kernels skip inactive IDs, preserving message and state offsets throughout Relay-BP. Before the OSD-0 phase, we compact unresolved sample IDs into a contiguous fallback queue. The solver uses these IDs to access message, posterior, and syndrome arrays at their original offsets, avoiding payload movement. We bound OSD-0 workspace through chunked execution because per-sample basis storage grows approximately quadratically with the number of checks. A conditional \texttt{\small{IF}} node executes each chunk only when the unresolved count exceeds that chunk's starting queue index. Kernels compute the unresolved count and evaluate all conditional nodes; the host stages inputs and launches the graph. For concurrent execution, each decoder instance owns a CUDA stream, private buffers, and an instantiated graph. Independent batches can advance through Relay legs and OSD-0 at different rates, exposing kernels from multiple decoder instances to the GPU scheduler.

\noindent\textbf{Warp-Level Collectives.} We parallelize neighborhood reductions and binary-vector operations within each syndrome sample using cooperating warp lanes. Cooperative check variants distribute each check-sample pair across a group of lanes and use shuffle reductions to combine sign parity and the two smallest magnitudes. In OSD-0, lanes update different packed words of a candidate vector and use shuffle reductions to select pivots. These operations exchange intermediate values through registers, avoiding shared-memory staging for the reductions.

\noindent\textbf{Ordered Multi-Warp OSD-0.} We further optimize OSD-0 by parallelizing reliability sorting and overlapping elimination across candidate variables. We use NVIDIA's \texttt{\small{CUB}} thread-block radix sort to order runs of 2,048 variables, then merge the sorted runs hierarchically. Merge-path partitioning divides each merge into independent output ranges. Composite keys combine the absolute posterior with the variable ID, breaking reliability ties deterministically. We reuse scratch storage across sorting and merging. For elimination, we assign one 256-thread block to each unresolved sample and pack matrix columns into 64-bit words, so each XOR updates 64 binary entries. The bivariate bicycle workload used in this study requires thirteen words per column, limiting parallelism within a single candidate. For this reason, we assign eight consecutive candidates to the block's eight warps and reduce them concurrently against the same unchanged basis. We then revisit candidates in reliability order, reducing each against any basis vectors inserted earlier in the group and inserting it if independent. This preserves sequential OSD-0 basis selection while overlapping independent elimination work. We stop basis construction at the precomputed matrix rank, then express the residual syndrome in that basis to construct the correction.

\noindent\textbf{Reduced Precision, Mixed Precision, and Quantization.} As a final optimization, we develop reduced-precision FP16, mixed FP16-FP32, and INT8-quantized Relay-BP kernels to reduce message storage and traffic. Each iteration reads and writes both directed edge-message arrays. For the bivariate bicycle workload used in this study, with a batch of 128 samples, these arrays occupy about 178~MiB in FP32 and 45~MiB in INT8. Because each concurrent decoder instance maintains private messages, the fourfold reduction also lowers the memory cost of concurrency. We select storage and arithmetic widths according to the numerical requirements of each update~\cite{Higham2022MixedPrecision}. Check updates combine signs and minimum magnitudes, with outgoing messages bounded by clipping. Variable updates sum incoming messages and retain estimates across iterations, producing posterior magnitudes that can exceed the message clipping threshold. Our mixed-precision implementation stores messages and Relay memory in FP16 and accumulates and stores posterior beliefs in FP32. The register-cached variable kernel reuses converted FP32 messages for posterior accumulation and outgoing-message generation. Our FP16 implementation also stores posterior state and performs accumulation in FP16, using native half-precision arithmetic and
\texttt{\small{half2}} for aligned initialization.

\begin{figure}[t]
\centering
\resizebox{0.7\columnwidth}{!}{%
\begin{tikzpicture}[
  font=\scriptsize,
  >=Latex,
  box/.style={
    draw,
    rounded corners=1.5pt,
    align=center,
    minimum height=7mm,
    inner sep=2.5pt
  },
  msg/.style={
    box,
    text width=23mm
  },
  op/.style={
    box,
    text width=34mm
  }
]

\node[msg] (channel) at (-1.65,0)
    {Channel prior\\\textbf{INT8}};

\node[msg] (c2v) at (1.65,0)
    {Check$\rightarrow$Variable\\\textbf{INT8}};

\node[
    op,
    text width=45mm,
    minimum height=16mm
] (belief) at (0,-1.65)
{
    \textbf{Belief/Relay update}\\[1.5pt]
    Accumulation: \textbf{INT32}\\
    Posterior: \textbf{INT16}\\
    Relay memory: \textbf{INT16}
};

\node[op] (quant) at (0,-3.10)
    {Scale + clip\\Quantize to \textbf{INT8}};

\node[msg] (v2c) at (0,-4.20)
    {Variable$\rightarrow$Check\\\textbf{INT8}};

\node[op] (check) at (0,-5.35)
    {\textbf{Parity/min-sum update}};

\draw[->]
    (channel.south)
    -- ++(0,-0.25)
    -| ([xshift=-9mm]belief.north);

\draw[->]
    (c2v.south)
    -- ++(0,-0.25)
    -| ([xshift=9mm]belief.north);

\draw[->] (belief.south) -- (quant.north);
\draw[->] (quant.south) -- (v2c.north);
\draw[->] (v2c.south) -- (check.north);

\draw[->]
    (check.east)
    -- ++(2.55,0)
    |- (c2v.east);

\end{tikzpicture}%
}
\caption{Relay-BP quantization with INT8 messages, INT32 accumulation, and INT16 posterior beliefs and Relay memory.}
\label{fig:int8_precision_flow}
\end{figure}
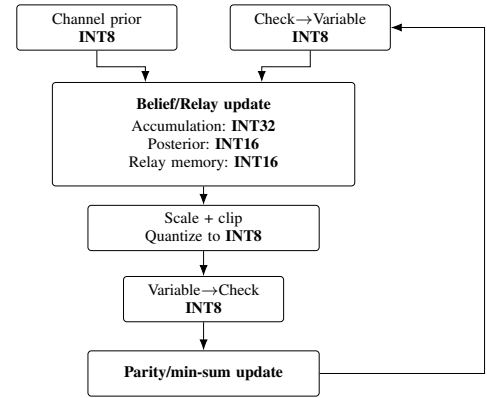

Fig.~\ref{fig:int8_precision_flow} distinguishes message storage, accumulation, and retained state in the INT8 implementation. We encode fractional reliabilities by multiplying them by six and rounding to an integer; for example, a reliability of 0.5 is stored as 3. Initial rounding selects the nearest integer, with ties away from zero. Clipping reliabilities at magnitude 20 gives the symmetric encoded range $[-120,120]$, simplifying sign and absolute-value operations. We apply normalized min-sum scaling with wider integer intermediates before rounding outgoing messages back to INT8. For variable updates, we accumulate incoming messages in INT32 and store posterior beliefs and Relay memory in INT16 at the same scale.
We implement the Relay blend with fixed-point coefficients and INT64 intermediate products. The wider state retains posterior magnitudes beyond the edge-message clipping threshold, preserving information used by OSD-0 to order variables by reliability. We apply clipping and conversion at the storage boundaries of these updates.

\noindent\textbf{Additional Optimization Experiments.} We also evaluated several kernel optimizations that were not retained in the final decoder. These included a two-samples-per-thread mapping with packed \texttt{\small{half2}} message accesses, four-message INT8 packing with \texttt{\small{uchar4}}, and 32-bit packed vectors for binary reduction. The packed-access variants targeted load/store instruction overhead. For binary post-processing, we separately tested Tensor Core 1-bit matrix multiply-accumulate instructions on dense binary tile products.

\begin{table}[t]
\centering
\caption{Circuit-level QEC workloads and decoder graph sizes.}
\label{tab:qec-workloads}
\scriptsize
\setlength{\tabcolsep}{3.0pt}
\renewcommand{\arraystretch}{1.10}

\begin{tabular}{lccc}
\toprule

\rowcolor{black!7}
\textbf{Parameter} &
\textbf{Surface} &
\textbf{HGP} &
\textbf{BB90} \\

\midrule

Code
& $[[81,1,9]]$
& $[[98,2,7]]$
& $[[90,8,10]]$ \\

Data qubits
& 81
& 98
& 90 \\

Stabilizer checks
& 80
& 98
& 90 \\

Syndrome-extraction rounds
& 9
& 9
& 9 \\

Two-qubit layers / round
& 8
& 8
& 12 \\

Detector checks
& 720
& 882
& 810 \\

DEM variables
& 12,391
& 18,522
& 30,869 \\

Tanner-graph edges
& 40,392
& 67,978
& 182,410 \\

Samples / noise point
& 10,000
& 10,000
& 10,000 \\

\bottomrule
\end{tabular}
\end{table}
\section{Experimental Setup}
\label{sec:experimental-setup}
\noindent\textbf{Workloads and Accuracy.}
We evaluate a rotated surface code, a hypergraph product (HGP) code~\cite{Tillich2014HGP}, and a bivariate bicycle code (BB90)~\cite{Bravyi2024BB}. Table~\ref{tab:qec-workloads} lists their circuits and decoding graphs. Their different graph sizes and sparsity patterns let us assess GPU performance across workloads with different message footprints, neighborhood reduction lengths, and indexed memory-access patterns. In $[[n,k,d]]$, $n$ counts physical data qubits, $k$ encoded logical qubits, and $d$ is the minimum weight of an undetected logical error. Each Z-memory experiment prepares a known encoded state, repeatedly measures stabilizer checks using auxiliary qubits (ancillas), and reads out in the computational basis. Deterministic edge colorings schedule the two check types separately. The noise model applies single-qubit depolarization before each round, two-qubit depolarization after controlled-NOT gates, post-reset flips, and measurement-result flips, all with $p\in\{0.001,0.003,0.005\}$. \texttt{\small{Stim}}~\cite{Gidney2021Stim} generates flattened DEMs retaining multi-detector errors and the samples listed in Table~\ref{tab:qec-workloads}. Identical samples are used across precisions and GPUs. Logical verification compares the predicted flip of one selected logical $Z$ observable, a parity of final readout bits, with Stim's sampled value. The \emph{logical error rate} (LER) is the fraction of samples with a mismatch after complete Relay-BP/OSD-0 decoding. 

\noindent\textbf{Decoder Configuration and Timing.} On the GPUs in Table~\ref{tab:gpu_platforms}, Relay-BP uses at most four legs of 32 iterations, a fixed memory coefficient of 0.5 across variables and legs, normalized min-sum with factor 0.75,
and message clipping at magnitude 20. Amortized decoding time is the synchronized decoder-call time divided by batch size $B$. Timing includes graph launch, Relay-BP, compaction, OSD-0, and internal synchronization; setup, transfers, and verification are excluded. For each GPU-staged batch, a host
monotonic timer records five repetitions after three warmups. Each GPU/workload/noise/precision configuration is independently
tuned over $B\in\{32,64,128,256,512,1024,2048\}$ using 2048 samples, two warmups, and three repetitions per candidate. Time reductions are relative to independently tuned FP32 on the same GPU, workload, and noise point.
\begin{table}[t]
\centering
\caption{Profiled GPU platforms.}
\label{tab:gpu_platforms}
\scriptsize
\setlength{\tabcolsep}{3.0pt}
\renewcommand{\arraystretch}{1.10}

\begin{tabular}{lccc}
\toprule

\rowcolor{black!7}
&
\multicolumn{2}{c}{\textbf{Data-center / HPC}} &
\textbf{Commodity} \\

\rowcolor{black!7}
\textbf{Parameter} &
\textbf{A100} &
\textbf{GH200} &
\textbf{RTX 5080} \\

Architecture
& Ampere
& Hopper
& Blackwell \\

Compute capability
& 8.0
& 9.0
& 12.0 \\

SMs
& 108
& 132
& 84 \\

CUDA-visible memory [GiB]
& 39.4
& 95.0
& 15.5 \\

Memory bus [bit]
& 5120
& 6144
& 256 \\

L2 cache [MiB]
& 40
& 60
& 64 \\

Max. GPU clock [GHz]
& 1.41
& 1.98
& 2.70 \\

Max. threads / SM
& 2048
& 2048
& 1536 \\

Max. warps / SM
& 64
& 64
& 48 \\

Shared memory / SM [KiB]
& 164
& 228
& 100 \\

Registers / SM
& 65,536
& 65,536
& 65,536 \\

ECC
& Yes
& Yes
& No \\

\bottomrule
\end{tabular}
\end{table}

\noindent\textbf{Decoding under Burst Load.} Synchronized QEC measurements can make multiple decoding requests available at once, motivating an evaluation of throughput and batch response time under burst load. To mimic this scenario, we design a simple benchmark. At $p=0.003$, we submit finite bursts of 32 independent batches to up to $S$ decoder instances, each with private buffers and a CUDA stream. An instance receives another batch upon completion. Requests include pinned-host syndrome upload, graph execution, and correction download. Batch response spans host submission to observed download completion, excluding time waiting for an available instance. Burst throughput spans first submission to final completion. We sweep $B\in\{16,32,64,128,256\}$ and S=1-4 on A100/GH200 or S=1-8 on RTX~5080. After four warmup batches, we measure five bursts and report the median of 160 batch responses and the median burst throughput. Precision is mixed FP16-FP32 for A100 Surface/HGP and RTX Surface, and INT8 for the remaining GPU/workload combinations.
\noindent\textbf{Kernel Profiling.} 
Nsight Compute profiles BB90 at $p=0.003$ in INT8 using each GPU's selected batch size. SM and DRAM SOL (``Speed of Light'') express streaming-multiprocessor and device-memory throughput relative to sustained peak. Active warps measure occupancy; eligible warps count ready-to-issue warps per scheduler in an active cycle. Issue activity is the percentage of scheduler cycles issuing an instruction. Long scoreboard reports warp stall cycles waiting for memory results from the L1/texture path, normalized by issue-active cycles. DRAM BW, L1/L2, and Regs/thread report device-memory bandwidth, cache hit rates, and per-thread register allocation, respectively.

\begin{table*}[t]
\centering
\caption{Amortized decoding time per sample. Time reductions and LER changes are relative to FP32.}
\label{tab:precision_results}
\scriptsize
\setlength{\tabcolsep}{3.0pt}
\renewcommand{\arraystretch}{1.10}

\resizebox{\textwidth}{!}{%
\begin{tabular}{lc|lrrrrl|lrrrrl||lrrrrl}
\toprule

\rowcolor{black!7}
& &
\multicolumn{6}{c|}{\textbf{A100}} &
\multicolumn{6}{c||}{\textbf{GH200}} &
\multicolumn{6}{c}{\textbf{RTX 5080}} \\

\rowcolor{black!7}
Code &
$p$ &
Prec. & $B$ &
\shortstack{Time\\($\mu$s)} &
\shortstack{Red.\\(\%)} &
\shortstack{OSD\\(\%)} &
\shortstack{LER (\%)\\$\Delta$ (pp)} &
Prec. & $B$ &
\shortstack{Time\\($\mu$s)} &
\shortstack{Red.\\(\%)} &
\shortstack{OSD\\(\%)} &
\shortstack{LER (\%)\\$\Delta$ (pp)} &
Prec. & $B$ &
\shortstack{Time\\($\mu$s)} &
\shortstack{Red.\\(\%)} &
\shortstack{OSD\\(\%)} &
\shortstack{LER (\%)\\$\Delta$ (pp)}
\\

\multirow{3}{*}{Surface}
& 0.001
& Mixed & 2048 & \favorable{72.786} & 10.0 & 1.18 & 0.01 ($+0.00$)
& INT8  & 2048 & \favorable{27.312} & 10.5 & 1.17 & 0.01 ($+0.00$)
& INT8  & 512  & \favorable{35.227} & 26.0 & 1.17 & 0.01 ($+0.00$)
\\

& 0.003
& Mixed & 2048 & 112.861 & 19.2 & 9.72 & 0.12 ($+0.00$)
& INT8  & 2048 & 40.812 & 26.8 & 9.85 & 0.14 ($+0.02$)
& Mixed & 256  & 51.326 & 27.7 & 9.72 & 0.12 ($+0.00$)
\\

& 0.005
& Mixed & 2048 & 121.984 & 25.7 & 28.07 & 1.40 ($-0.03$)
& INT8  & 2048 & 41.927 & 34.1 & 28.10 & 1.49 ($+0.06$)
& INT8  & 256  & 53.936 & 29.4 & 28.10 & 1.49 ($+0.06$)
\\

\midrule

\multirow{3}{*}{HGP}
& 0.001
& Mixed & 2048 & 107.166 & 3.8 & 1.08 & 0.00 ($+0.00$)
& FP16  & 2048 & 38.012 & 10.1 & 0.68 & 0.00 ($+0.00$)
& INT8  & 256  & 58.217 & 9.9 & 1.28 & 0.00 ($+0.00$)
\\

& 0.003
& Mixed & 2048 & 181.258 & 18.8 & 9.68 & 0.07 ($+0.00$)
& INT8  & 2048 & 62.427 & 29.9 & 9.86 & 0.08 ($+0.01$)
& INT8  & 256  & 79.906 & 43.5 & 9.86 & 0.08 ($+0.01$)
\\

& 0.005
& Mixed & 2048 & 198.378 & 26.8 & 31.03 & 0.43 ($-0.02$)
& INT8  & 2048 & 65.411 & 36.5 & 30.54 & 0.52 ($+0.07$)
& INT8  & 256  & 87.452 & 41.8 & 30.54 & 0.52 ($+0.07$)
\\

\midrule

\multirow{3}{*}{BB90}
& 0.001
& Mixed & 1024 & 145.097 & 17.2 & 0.07 & 0.00 ($+0.00$)
& Mixed & 1024 & 55.537 & 14.7 & 0.07 & 0.00 ($+0.00$)
& INT8  & 128  & 52.501 & 34.9 & 0.09 & 0.00 ($+0.00$)
\\

& 0.003
& INT8 & 2048 & 423.136 & 15.4 & 4.73 & 0.65 ($+0.03$)
& INT8 & 2048 & 151.154 & 22.6 & 4.73 & 0.65 ($+0.03$)
& INT8 & 128  & \favorable{207.079} & \favorable{50.2} & 4.73 & 0.65 ($+0.03$)
\\

& 0.005
& INT8 & 2048 & 467.740 & 31.2 & \tradeoff{30.85} & \tradeoff{5.51 ($+0.32$)}
& INT8 & 2048 & 165.692 & 34.8 & \tradeoff{30.85} & \tradeoff{5.51 ($+0.32$)}
& INT8 & 128  & \favorable{236.442} & \favorable{55.6} & \tradeoff{30.85} & \tradeoff{5.51 ($+0.32$)}
\\

\bottomrule
\end{tabular}%
}
\end{table*}

\section{Results}
\label{sec:results}
We present nine findings covering decoder performance, GPU resource use, and concurrent batch processing. 

\keyresult{1}{Batch-size autotuning is essential to single-decoder performance.} Batch size determines how much independent work the GPU can schedule. Larger batches amortize graph-launch and control overheads, supply more independent warps to hide memory and synchronization stalls, and expose
more concurrent OSD solvers. This is especially important after sample retirement. For BB90 in mixed precision on A100 at $p=0.003$, increasing the batch size from 32 to 2048 reduces amortized decoding time from 1466 to 439~$\mu$s/sample, a $3.34\times$ speedup. The optimum also changes with noise: at $p=0.001$, the selected batch is 1024, and doubling it to 2048 increases amortized time by 10.3\%. 

\keyresult{2}{CUDA Graphs unify execution with limited timing gains.} In this work, CUDA Graphs express iteration control, sample retirement, and selective fallback within a reusable device-controlled workflow. We find that conditional control provides little timing benefit when unresolved samples keep each batch active through the iteration limit. For BB90 on A100 in FP32 with $B=256$, static and conditional graphs take approximately 607 and 614~$\mu$s/sample at $p=0.003$. At $p=0.001$, conditional execution reduces time from 229 to 191~$\mu$s/sample by terminating completed batches earlier. The benefit depends on whole-batch convergence. The main CUDA Graph contribution is a unified execution path with device-side control. 

\keyresult{3}{Reduced precision, mixed precision, and quantization are key to efficient GPU QEC decoding.} Table~\ref{tab:precision_results} shows complete-decoder timing gains across all three GPUs. The fastest format depends on architecture and workload: A100 selects mixed precision for Surface and HGP, while INT8 is selected more often on GH200 and RTX~5080. Narrower messages reduce storage and traffic, but conversion costs and changes in convergence determine the complete-decoder benefit. In Table~\ref{tab:precision_results}, the bold green Surface entries at $p=0.001$ mark the shortest Surface decoding times on each GPU, with unchanged reported LER. The green RTX~5080 entries for BB90 highlight larger relative gains. At $p=0.003$, INT8 reduces amortized time from 416 to 207~$\mu$s/sample, approximately doubling performance, while observed LER changes from 0.62\% to 0.65\%. At $p=0.005$, the time reduction reaches 55.6\%. The bold red entries identify the accompanying higher-noise costs. For BB90 at $p=0.005$, INT8 increases observed LER from 5.19\% to 5.51\% relative to FP32. Its OSD fallback fraction reaches 30.85\%, compared with 4.73\% at $p=0.003$, shifting more work into sorting and binary elimination. 

\keyresult{4}{Warp-level collectives accelerate QEC decoding.} Warp-level collectives expose parallelism inside the small reductions and binary-vector updates that recur throughout QEC. Register exchanges between lanes reduce serial work and shared-memory staging for local reductions, complementing concurrency across independent syndromes.

OSD-0 demonstrates the impact at both kernel and decoder levels. On A100, a standalone benchmark with 810 checks and 640 synthetic residuals achieves 20-21$\times$ faster packed reduction than a one-thread-per-sample scalar reference. Extending the production warp-cooperative solver to eight concurrent candidate reductions gives a further 2.25$\times$ solver-kernel speedup. Candidates are finalized in reliability order, incorporating preceding basis updates and preserving sequential OSD-0 selection. For BB90 at p=0.005, FP32, and B=256, complete-decoder time decreases by 8.8\%, with unchanged logical-error counts. Profiling shows less barrier waiting and higher scheduler issue activity.

The cooperative INT8 BP phase applies the same principle to message passing. Lane groups partition each check neighborhood and use warp shuffles to combine sign parity and the two minimum message magnitudes before computing outgoing messages. 

\keyresult{5}{Binary Tensor Core instructions provide modest gains in binary algebra.} QEC post-processing relies on binary linear algebra, making Tensor Cores an interesting acceleration target. NVIDIA's 1-bit matrix multiply-accumulate instructions process packed binary tiles using bitwise AND and population counts. Extracting each count's parity produces the binary matrix product. On A100, we compare this Tensor Core path with packed AND and
population-count operations. For 65,536 products with $8\times8$ output tiles and inner dimension 128, kernel time decreases from 30.77 to 27.43~$\mu$s, a 1.12$\times$ speedup with matching outputs. This modest gain is measured on regular matrix products with inputs already packed. Exploiting Tensor Cores within the OSD-0 solver would require blocked binary updates that preserve these dependencies and amortize tile preparation. 

\keyresult{6}{L2 reuse and DRAM utilization differ across tuned GPU configurations.} Table~\ref{tab:gpu_profile} compares BB90 kernel profiles at each
GPU's selected INT8 batch size. For each kernel, L1 hit rates remain similar across devices; the larger memory-system differences appear in L2 reuse and DRAM utilization. The green \texttt{\small{Belief/Relay}} L2 entries on A100 and GH200 show 74\% and 65\% hit rates, compared with the red RTX~5080 entry at 36\%. The bold DRAM entries distinguish relative utilization from absolute bandwidth. RTX~5080 reaches 55\% DRAM SOL while transferring 0.52~TB/s; GH200 transfers 1.13~TB/s at 28\% SOL. The selected RTX configuration consumes a larger fraction of its available bandwidth, while GH200 sustains more than twice the transfer rate at lower relative DRAM utilization.

\begin{table}[t]
\centering
\caption{Relay-BP kernel profiles for BB90 at $p=0.003$ (INT8), with batch sizes from Table~\ref{tab:precision_results}.}
\label{tab:gpu_profile}
\scriptsize
\setlength{\tabcolsep}{3.0pt}
\renewcommand{\arraystretch}{1.10}

\begin{tabular}{
    >{\raggedright\arraybackslash}p{0.34\columnwidth}
    >{\centering\arraybackslash}p{0.27\columnwidth}
    >{\centering\arraybackslash}p{0.27\columnwidth}}
\toprule

\rowcolor{black!7}
\textbf{Metric}
& \texttt{Parity/min-sum} kernel
& \texttt{Belief/Relay} kernel \\

\midrule

\rowcolor{black!7}
\multicolumn{3}{l}{\textbf{A100} \hfill $B=2048$} \\

SM SOL [\%]
& 62.21 & 64.98 \\

DRAM SOL [\%]
& 24.77 & 24.93 \\

DRAM BW [TB/s]
& 0.385 & 0.388 \\

Active warps [\%]
& 96.54 & 73.45 \\

Issue active [\%]
& \tradeoff{51.38} & \favorable{65.16} \\

L1 hit [\%]
& 25.20 & 74.71 \\

L2 hit [\%]
& 65.65 & \favorable{73.92} \\

Eligible warps / scheduler
& \tradeoff{1.32} & \favorable{2.05} \\

Long scoreboard [ratio]
& \tradeoff{23.77} & \favorable{9.61} \\

Regs/thread
& 32 & 40 \\

\midrule

\rowcolor{black!7}
\multicolumn{3}{l}{\textbf{GH200} \hfill $B=2048$} \\

SM SOL [\%]
& 55.58 & 63.23 \\

DRAM SOL [\%]
& 24.60 & \textbf{28.19} \\

DRAM BW [TB/s]
& 0.990 & \textbf{1.134} \\

Active warps [\%]
& 71.84 & 73.63 \\

Issue active [\%]
& 46.73 & 63.49 \\

L1 hit [\%]
& 24.62 & 72.46 \\

L2 hit [\%]
& 55.07 & \favorable{64.89} \\

Eligible warps / scheduler
& 0.90 & 1.71 \\

Long scoreboard [ratio]
& 19.93 & 11.42 \\

Regs/thread
& 36 & 40 \\

\midrule

\rowcolor{black!7}
\multicolumn{3}{l}{\textbf{RTX 5080} \hfill $B=128$} \\

SM SOL [\%]
& 40.99 & 47.90 \\

DRAM SOL [\%]
& 34.63 & \textbf{54.96} \\

DRAM BW [TB/s]
& 0.328 & \textbf{0.520} \\

Active warps [\%]
& 68.00 & 89.44 \\

Issue active [\%]
& 38.59 & 48.78 \\

L1 hit [\%]
& 25.56 & 74.41 \\

L2 hit [\%]
& 58.09 & \tradeoff{35.87} \\

Eligible warps / scheduler
& 0.65 & 1.15 \\

Long scoreboard [ratio]
& 16.42 & 15.06 \\

Regs/thread
& 40 & 40 \\

\bottomrule
\end{tabular}
\end{table}

\keyresult{7}{Memory dependencies limit instruction issue at high occupancy.} On A100, \texttt{\small{Parity/min-sum}} reaches 97\% occupancy, yet records 51\% issue activity and 1.32 eligible warps per scheduler (both highlighted in red). \texttt{\small{Belief/Relay}} achieves 65\% issue activity and 2.05 eligible warps per scheduler (green) at 73\% occupancy. Across all three GPUs, \texttt{\small{Belief/Relay}} consistently supplies more eligible warps and achieves higher issue activity than \texttt{\small{Parity/min-sum}}. The highlighted long-scoreboard ratios connect this readiness gap
to memory dependencies: A100 records 23.77 warp stall cycles per issue-active cycle for the \texttt{\small{Parity/min-sum}} kernel (red), compared with 9.61 for the \texttt{\small{Belief/Relay}} kernel (green). Resident warps waiting for memory operands cannot advance their dependent instructions. 

\begin{figure}[t]
\centering
\includegraphics[width=\columnwidth]{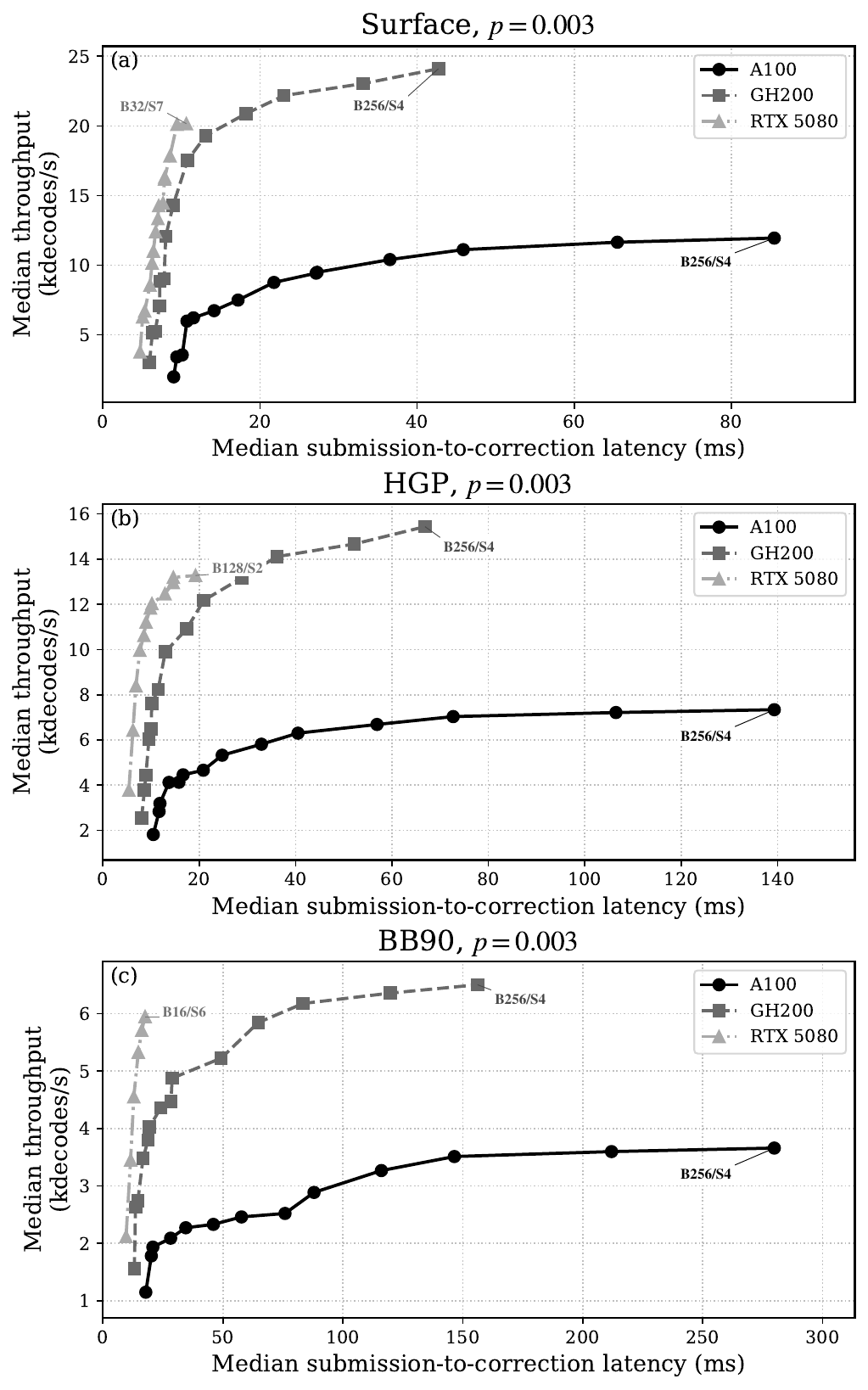}
\caption{Latency-throughput Pareto frontiers at $p=0.003$. Endpoint annotations give batch size $B$ and stream count $S$ at the highest-throughput point.}
\label{fig:pareto_frontiers}
\end{figure}

\keyresult{8}{Batch size and stream concurrency control the latency-throughput tradeoff.} Fig.~\ref{fig:pareto_frontiers} identifies which combinations of batch size and stream count deliver the highest measured throughput within a given response-time constraint under burst input. The Surface workload on GH200 illustrates the benefit of concurrency at small batch sizes. At $B=16$, increasing the stream count from one to four raises throughput from 3.01 to 9.02~kdecodes/s, a $3\times$ gain, while median response increases from 5.93 to 7.78~ms. Concurrent batches supply more independent work while preserving
the small batch size of each request. Larger batches show diminishing throughput returns near the frontier's upper endpoint. At $S=4$, doubling $B$ from 128 to 256 increases throughput from 22.18 to 24.10~kdecodes/s, an 8.7\% gain, while response increases from 23.05 to 42.72~ms, an 85\% increase.
For a 10~ms median-response requirement, the best measured configuration is $B=32$, $S=4$, delivering 14.30~kdecodes/s at 8.94~ms. 

\keyresult{9}{Commodity GPUs can outperform HPC GPUs within a response-time requirement.} Comparing the frontiers under fixed response-time limits changes the GPU ranking. Across all three workloads, every A100 frontier point has an RTX~5080 configuration delivering higher throughput with a shorter
response. RTX also exceeds GH200's throughput under short response-time limits. For Surface within 10~ms, RTX delivers 20.11~kdecodes/s compared
with GH200's 14.30~kdecodes/s, a 41\% advantage. For BB90 within 20~ms, the corresponding rates are 5.95 and 4.03~kdecodes/s, a 48\% advantage. GH200 reaches the highest measured throughput for all three code families, with larger batches extending its frontier toward higher throughput and longer responses. On BB90, GH200 delivers 6.50~kdecodes/s with a 156~ms median response; RTX delivers 5.95~kdecodes/s at 17.6~ms, providing 91\% of that throughput with an 8.9-times shorter response. RTX uses six concurrent decoder instances processing 16-sample batches, compared with GH200's four instances processing 256-sample batches.

RTX's efficiency at small batch sizes is key to this latency-throughput advantage. Small batches provide shorter units of work. Each completion returns corrections and releases a decoder instance for the next request, while other instances continue executing. Multiple instances supply parallelism across requests, allowing RTX to sustain high throughput with less work within each batch. A100 and GH200 benefit more from larger batches to reach high throughput, increasing the work each request must complete before returning corrections. For 32-bit integer addition and minimum instructions, NVIDIA lists 128 results per clock per SM on RTX Blackwell, compared with 64 on Hopper.\footnote{\href{https://docs.nvidia.com/cuda/archive/12.9.1/cuda-c-programming-guide/index.html\#arithmetic-instructions}
{NVIDIA CUDA Guide, Table~7}, accessed Oct.~20, 2026.} This higher instruction throughput provides a plausible additional advantage for the INT8 updates, which repeatedly accumulate message values and select minimum magnitudes.

\section{Related Work}
BP with OSD is an established decoder for quantum low-density parity-check codes~\cite{Roffe2020Landscape}. Relay-BP advances message-passing accuracy
through memory and relay ensembling~\cite{Mueller2025RelayBP}; its strong circuit-noise results motivate our choice of algorithmic foundation.
Localized statistics decoding parallelizes post-processing by using binary solvers on local regions of the decoding graph~\cite{Hillmann2025LSD}. B\'ascones et al.~\cite{Bascones2025FPGAASICBPOSD} study FPGA/ASIC implementations of BP and OSD, including elimination cost. On GPUs, Ferraz et al.~\cite{Ferraz2025GPU} implement BP syndrome decoding on commodity hardware. CUDA-Q QEC provides BP/OSD and Relay-BP
decoders~\cite{NVIDIACudaQQEC}, and Wang et al.~\cite{Wang2026GPUBP} investigate parallel BP retries and windowed/streaming execution. Maurya et al.~\cite{Maurya2026Elastic} study provisioning and scheduling of shared decoders under bursty demand. Knapen et al.~\cite{Knapen2026Predecoding} develop qLDPC hardware predecoders that reduce OSD invocation and additional Relay-BP legs.

\section{Conclusion}
We designed and implemented a Relay-BP/OSD-0 decoder entirely on NVIDIA GPUs, combining specialized kernels, convergence control with CUDA Graphs, ordered multi-warp binary elimination, reduced and mixed precision, and quantization. We evaluated decoding time, logical recovery, and latency-throughput tradeoffs for concurrent decoding requests across three circuit-level code families and three GPU architectures. Relatively small concurrent batches allow a commodity RTX GPU to approach GH200's peak burst throughput with substantially shorter responses.

These results support commodity GPU clusters as programmable HPC infrastructure for QEC, combining competitive decoding throughput with low-latency responses. Such clusters could become standard components of fault-tolerant quantum computers, serving as programmable decoding accelerators alongside quantum controllers.

\section*{Acknowledgment}
 {This work was supported by the Wallenberg Centre for Quantum Technology (WACQT), funded by the Knut and Alice Wallenberg Foundation (KAW).}

\balance
\bibliographystyle{IEEEtran}
\bibliography{QECprecision}
\end{document}